\documentclass[useAMS,usenatbib]{mn2e}
\usepackage{graphicx}
\usepackage[authoryear]{natbib}
\usepackage{longtable}

\title[The TF relation for HCG galaxies]{The Tully-Fisher relations for Hickson Compact Group galaxies\thanks{HCG Fabry-Perot data are available at http://fabryperot.oamp.fr}}
\author[S. Torres--Flores et al.]{S. Torres-Flores
$^{1}$\thanks{E-mail: storres@dfuls.cl}, C. Mendes de Oliveira$^{2}$, H. Plana$^{3}$, P. Amram$^{4}$, B. Epinat$^{4}$\\
$^{1}$Departamento de F\'isica, Universidad de La Serena, Av. Cisternas 1200 Norte, La Serena, Chile \\
$^{2}$Departamento de Astronomia, Instituto de Astronomia, Geof\'isica e Ci\^encias Atmosf\'ericas da USP,\\
Rua do Mat\~ao 1226, Cidade Universit\'aria, 05508-090, S\~ao Paulo, Brazil\\
$^{3}$Laborat\'orio de Astrof\'isica Te\'orica e Observacional, Universidade Estadual de Santa Cruz, Ilh\'eus, Brazil\\
$^{4}$Laboratoire d'Astrophysique de Marseille, Aix Marseille Universit\'e, CNRS, 13388, Marseille, France}

\begin{document}

\date{}

\pagerange{\pageref{firstpage}--\pageref{lastpage}} \pubyear{2012}

\maketitle

\label{firstpage}

\begin{abstract}
We used K-band photometry, maximum rotational velocities derived from Fabry-Perot data and HI observed and predicted masses to study, for the first time, the K-band, stellar and baryonic Tully-Fisher relations for galaxies in Hickson compact groups. We compared these relations with the ones defined for galaxies in less dense environments from the GHASP survey and from a sample of gas-rich galaxies. We find that most of the Hickson compact group galaxies lie on the K-band Tully-Fisher relation defined by field galaxies with a few low-mass outliers, namely HCG 49b and HCG 96c, which appear to have had strong recent burst of star formation. The stellar Tully-Fisher relation for compact group galaxies presents a similar dispersion to that of the K-band relation, and it has no significant outliers when a proper computation of the stellar mass is done for the strongly star-forming galaxies. The scatter in these relations can be reduced if the gaseous component is taken into account, i.e., if a baryonic Tully-Fisher relation is considered. In order to explain the positions of the galaxies off the K-band Tully-Fisher relation we favour a scenario in which their luminosities are brightened due to strong star
formation or AGN activity. We argue that strong bursts of star formation can affect the B and K-band luminosities of HCG 49b and HCG 96c and in the case of the latter also AGN activity may affect the K-band magnitude considerably, without affecting their total masses.
\end{abstract}

\begin{keywords}
galaxies: evolution, galaxies: interactions, galaxies: kinematics and dynamics
\end{keywords}

\section{Introduction}

There has been increased interest in the past years in the use of the
Tully-Fisher relation (TF) as a means of quantifying galaxy evolution
as a function of redshift. Given that it is well known that interactions
are much more frequent in the distant universe than at present, one must
understand the effects of the dense environments over the structure of
nearby galaxies before attempting to understand evolutionary effects as
a function of redshift (e. g. Kannappan et al. 2004).

A powerful way to analyze the Tully-Fisher relation is to take into
account not only the stellar mass but also the gaseous mass of the
galaxies and therefore to have an estimate of the total baryonic
mass. This relation has been studied by several authors (McGaugh 2000,
Bell et al. 2003b, Geha et al. 2006, Kassin et al. 2006, Torres-Flores et al. 2011, Foreman \& Scott 2011, Gnedin 2011, McGaugh 2011) and it has been shown to be very
important in the evolution of galaxies and in the determination of parameters such as the mass-to-light
ratio for late-type galaxies (Bell \& de Jong 2001).

In the last years, most of the studies of the baryonic Tully-Fisher
relation (BTF) have focussed on dwarf galaxies, gas dominated and
extremely low-mass systems (Geha et al. 2006, Stark et al. 2009 and Begum
et al. 2008 respectively). No comparison between the BTF for field and
interacting galaxies has been carried out. Whether tidal forces can remove some of the dark halo in interacting
galaxies (Rubin et al. 1991) is still an open question, especially when some works do not show large differences
in the dark matter distribution of interacting galaxies with respect to field galaxies (e. g. Plana et al. 2010).

Among the best places to study interacting galaxies are compact groups,
where tidal encounters are common, due to the low velocity dispersion of
the group and the high spatial density of the galaxies. Therefore, it is expected 
that interactions have disturbed the kinematics and/or the star formation rate of galaxies 
in compact groups. One of the first catalogues of compact groups was published by Hickson (HCG, 1982). This catalogue contains a sample 100 groups, which are in different evolutionary stages. These groups have been studied 
extensively, in order to understand the evolution of galaxies in dense environments. 

As shown in Mendes de Oliveira et al. (2003) and Torres-Flores et
al. (2010), most of the HCG galaxies lie on the B-band TF relation,
with a few low-mass outliers. The low-mass galaxies tend to present a low
rotational velocity for their observed luminosity or a high luminosity
for their observed rotational velocity. If included in the fit, these
low-mass galaxies affect strongly the determination of the slope and
zero-point of the TF relation.

In this paper we study for the first time the K-band, stellar and baryonic Tully-Fisher
relations for a sample of HCG galaxies and compare them with
the corresponding relations for galaxies in less dense environments. Here, 
we also attempt to find possible explanations for the
location of the outlying positions of the  low-mass galaxies in the
B-band compact group TF relation (Mendes de Oliveira et al. 2003 and
Torres-Flores et al. 2010). Studying the K-band, stellar and baryonic
Tully-Fisher relation we could investigate if the mass of the galaxies
could be truncated due to interactions or if a burst of star formation
could be strong enough to increase the luminosity of these objects,
pushing them off the TF-relations. In order to carry out this analysis,
we have used a sample of Hickson compact group galaxies for which rotation
curves and photometry are available (Mendes de Oliveira et al. 2003 and
Torres-Flores et al. 2010). As control sample we have used a sample of
non-interacting galaxies from the GHASP survey, for which the baryonic
Tully-Fisher relation has been recently studied by Torres-Flores et
al. (2011). These authors found that the baryonic TF relation for the GHASP
sample is in agreement with cosmological predictions (e. g. Bullock et
al. 2001). The GHASP sample has been complemented with a sample of gas-rich galaxies, which was published in McGaugh et al. (2012). This paper is organized as follows. In \S
2 we present the data. In section \S 3 we present our results. In \S 4
we discuss our results and in \S 5 we give our conclusions.

\section{Data}

\subsection{Sample}

The sample of Hickson compact group galaxies used in this work is composed by groups in the local universe, with radial velocities lower than 10500 km s$^{-1}$. These groups are in different evolutionary stages, containing strongly to mildly interacting systems. 

Rubin et al. (1991) have reported that only a third of the galaxies in compact group have rotation curves which are well behaved enough to be placed in the Tully-Fisher relation. In the same context, Torres-Flores et al. (2010) found that rotation curves can be derived for 68$\%$ of the observed HCG galaxies, suggesting that this fact could be associated with the complex kinematics of these systems. 

Furthermore, for this work, we have selected a sample of 36 HCG galaxies for which rotation curves are published (Mendes de Oliveira et al. 2003 and Torres-Flores et al. 2010). 

Among the galaxies chosen in this study, a few have asymmetric rotation curves (common in systems in interaction). Nevertheless, maximum velocities could still be derived, to place them in the Tully-Fisher relation. 

It is important to note, that the rotation curves published by Mendes de Oliveira et al. (2003) and Torres-Flores et al. (2010) were derived from 2D velocity fields. The use of 2D velocity fields in the construction of the rotation curves avoid any assumption about the major axis position angle and inclination of the galaxies as is the case for long-slit observations. For the galaxies taken from Mendes de Oliveira et al. (2003), maximum rotational velocities were defined as the maximum values  for the average velocities for the approaching and receding sides. For the galaxies taken from Torres-Flores et al. (2010), maximum rotational velocities were computed as defined in Epinat et al. (2008a).

Although the compact group galaxies studied here do not comprise a complete sample in magnitude, they are representative of the population of the most undisturbed compact group galaxies, and they span a similar range of absolute magnitudes and a similar mix of morphological types as the galaxies in the control sample. Despite its incompleteness, this sample allows a fair comparison of the positions of compact group and field galaxies in the Tully-Fisher relation and identification of outliers.

\subsection{Control sample}

The control sample used for this study is from the Gassendi HAlpha survey of Spirals (GHASP; Garrido et al. 2002, 2003, 2004, 2005, Epinat et al. 2008a,b). The GHASP survey is the largest sample of spiral and irregular galaxies observed with a Fabry-Perot instrument. It consists of 203 galaxies, covering a large range of absolute magnitudes. For all these galaxies, 3D H$\alpha$ data cubes were recently analyzed in a homogeneous way by Epinat et al. (2008a,b). These authors published velocity fields, rotation curves and maximum velocities (V$_{max}$) for all the galaxies.   

By using K-band photometry (from 2MASS) and the V$_{max}$ published by Epinat et al. (2008a,b), Torres-Flores et al. (2011) studied the near infrared, stellar and baryonic Tully-Fisher relation for 46 galaxies of the GHASP sample. This sub-sample was selected in order to have galaxies with inclinations higher than 25 degrees (to reduce the uncertainties on the rotational velocities), galaxies with accurate distance measurements and galaxies for which there were 2MASS K-band and \textit{Sloan Digital sky Survey photometry} (\textit{SDSS}) photometry (the latter was required in order to estimate the mass-to-light ratio of the galaxies). In the end, this GHASP sub-sample covers a mass range of $8.91\leq\textrm{log}(M_{baryonic}/M_{\odot})\leq11.37$. Torres-Flores et al. (2011) found that the slope of the baryonic Tully-Fisher relation for the GHASP sample is in agreement with those determined by previous works (e. g. Geha et al. 2006).

As we are interested in comparing the near infrared, stellar and baryonic Tully-Fisher relations for galaxies in different environments, it is important to note that, for both samples, velocity fields are available for all objects and the samples were analyzed in the same way, using the same procedures to obtain the rotation curves and the maximum rotational velocity of each galaxy. These facts make the GHASP sub-sample a suited sample for comparing the HCG Tully-Fisher relation and allows us to perform a fair comparison between both samples. In this sense, systematic errors should not affect our main results.

\begin{figure}
\includegraphics[width=\columnwidth]{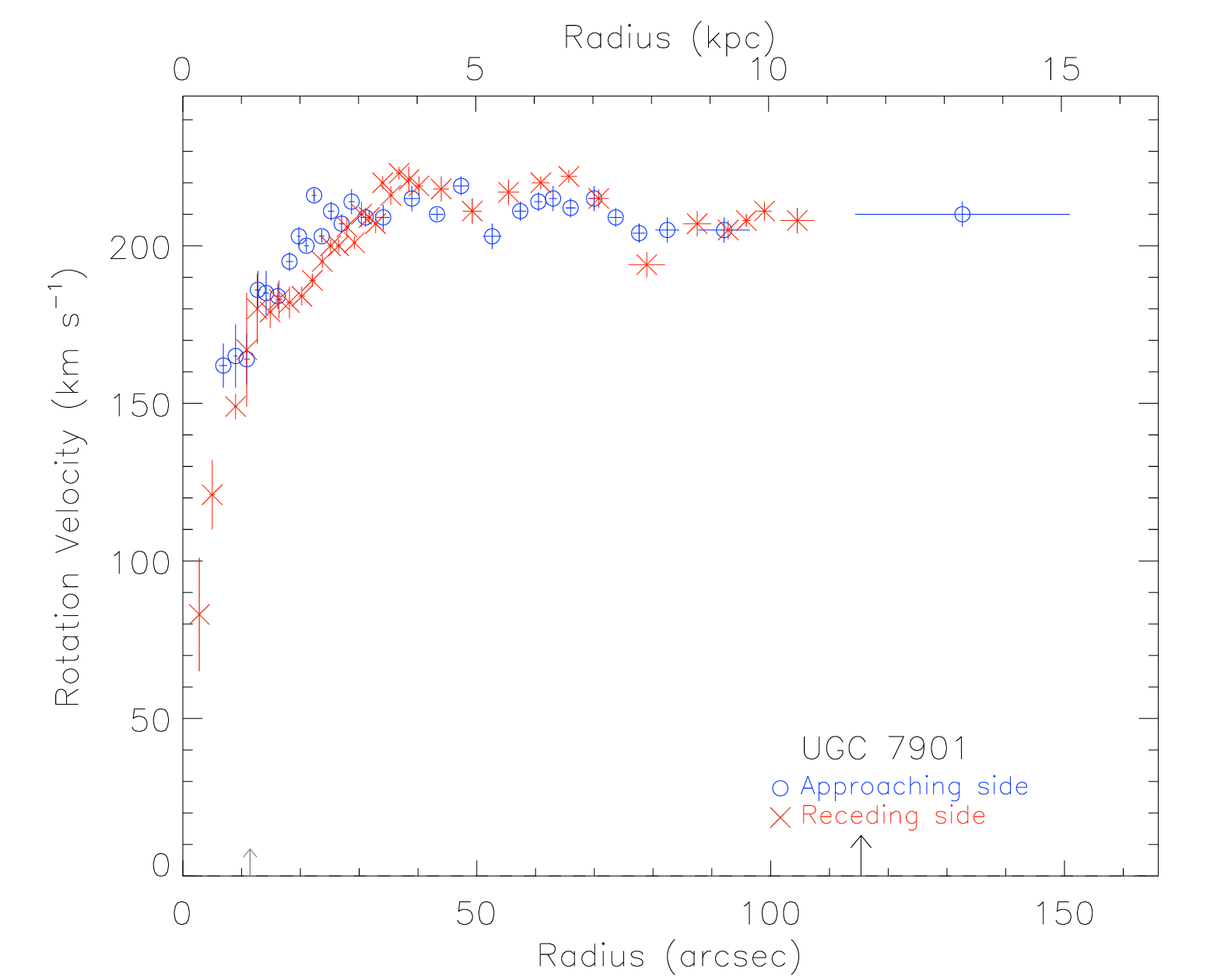}
\includegraphics[width=\columnwidth]{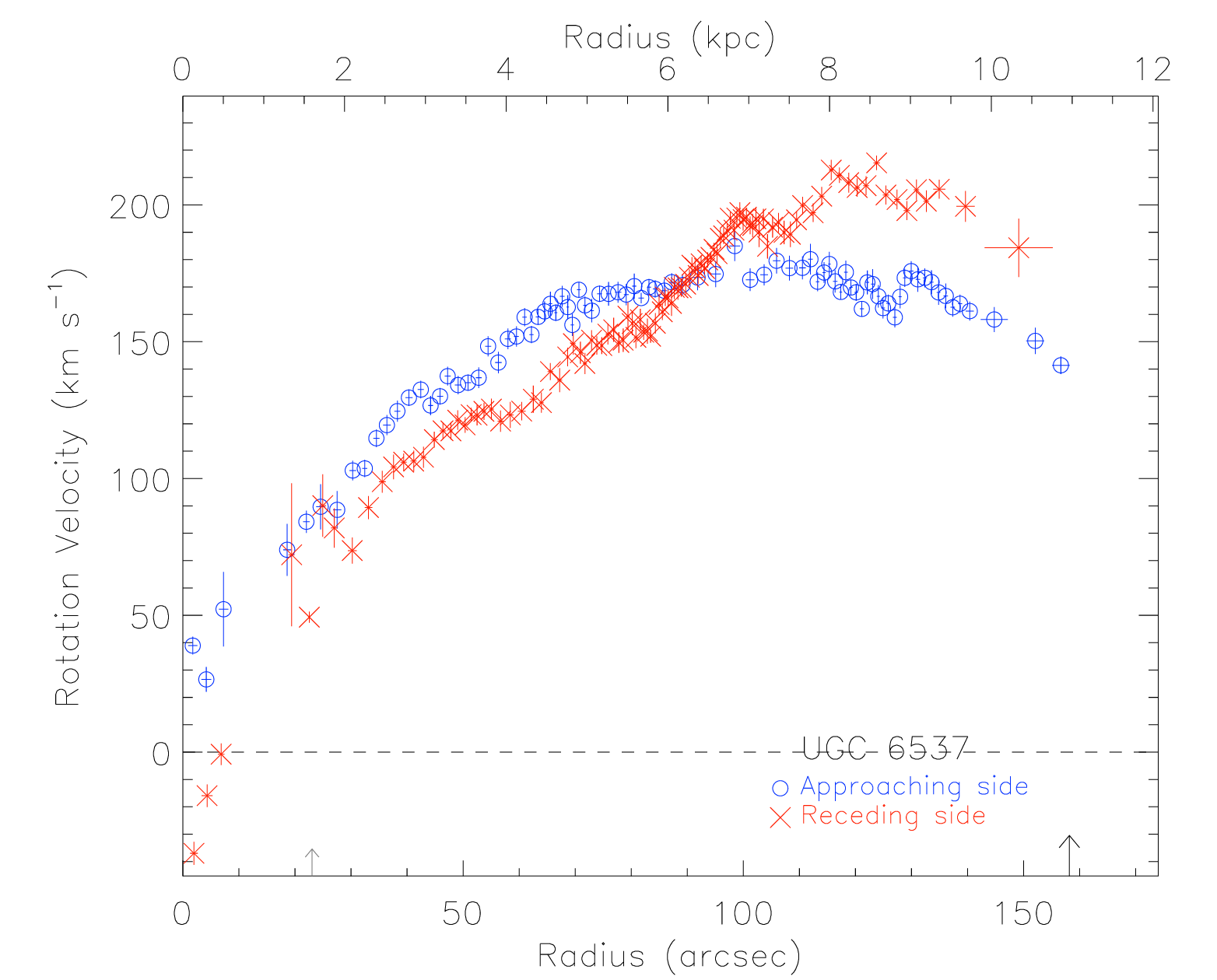}
\caption{Top Panel: Rotation curve for UGC 7901, which shows a symmetric behaviour. Bottom panel: Rotation curve for UGC 6537. This curve displays a clear asymmetry between the receding and approaching sides.}
\label{fig0}
\end{figure}

Torres-Flores et al. (2011) divided the GHASP sample between galaxies displaying symmetric, slightly asymmetric and asymmetric rotation curves. In this paper, we have included the slightly asymmetric and asymmetric rotation curves in just one class. Therefore, in the comparisons of the Tully-Fisher relations, the GHASP sample will be divided in symmetric and asymmetric rotation curves. As an example, in Fig. \ref{fig0} we show a symmetric and an asymmetric rotation curve coming from the GHASP sample, where the figures were taken from Epinat et al. (2008) and Epinat et al. (2008b) (top and bottom panel respectively).

Given that we are interested in the location of the low-mass HCG galaxies on the Tully-Fisher relation defined by galaxies in less dense environments, we have complemented the GHASP sample with the gas-rich galaxies studied by McGaugh (2012), who takes this sample from the previous works published by Begum et al. (2008), Stark et al. (2009) and Trachternach et al. (2009). As McGaugh (2012) lists the stellar and gaseous mass of the gas-rich sample, we will use it in the comparison of the stellar and baryonic Tully-Fisher relations.

\subsection{Photometry}

We used the K-band magnitudes within the isophote of 20 mag arcsec$^{-1}$ (m$_{K20}$) from 2MASS (Skrutskie et al. 2006) for the HCG sample, given that this band is more reflective of the stellar content of galaxies and is less affected by dust and ongoing star formation than optical bands. K-band magnitudes were corrected for Galactic extinction ($A_{K}$) using the Schlegel maps (Schlegel et al. 1998). Extinction corrections due to the inclinations were applied using the method given in Masters et al. (2003). k-corrections ($k_{K}$) were also taken from that study. Absolute magnitudes were obtained using:

\begin{equation}
 M_{K}=m_{K20}-5\times log(D)+k_{K}-A_{K}-25
 \label{mk} 
\end{equation}
 
where $D=vvir/H_{0}$. Here, vvir corresponds to the heliocentric radial velocity of the galaxy, corrected for Local Group infall onto Virgo, from Hyperleda and H$_{0}$=75 km s$^{-1}$ Mpc$^{-1}$. For HCG 2a, we measured the K-band magnitude within the isophote of 20 mag arcsec$^{-1}$ using the task ELLIPSE in IRAF\footnote{IRAF is distributed by the National Optical Astronomy Observatories, which are operated by the Association of Universities for Research in Astronomy, Inc., under cooperative agreement with the National Science Foundation.}, finding that this object should be 0.77 magnitudes brighter than the value given in 2MASS. We also perform this comparison for other galaxies (in particular for HCG 49b and HCG 96c) and the values obtained using ELLIPSE were in perfect agreement with those given by 2MASS. In the following, we will use the K-band magnitude of HCG 2a obtained from our analysis. For all other galaxies m$_{K20}$, from 2MASS, was used. Columns 1 and 2 in Table \ref{table1} list the name and K-band absolute magnitudes for the HCG galaxies.

Finally, K-band luminosities were estimated using 

\begin{equation}
L_{K}=10^{-0.4(M_{K}-3.41)} 
\label{lk}
\end{equation}

In the case of B-band luminosity, it was derived using:

\begin{equation}
L_{B}=10^{-0.4(M_{B}-5.49)} 
\label{lb}
\end{equation}

where M$_{B}$ was estimated as in Torres-Flores et al. (2010).

\subsection{Fitting method}

In order to do a fair comparison of the Tully-Fisher relation for HCG and field galaxies, we have used the same fitting procedure as Torres-Flores et al. (2011) for the GHASP sample. Basically, we perform a linear fit to the observed data, taking into account the uncertainties in the magnitudes (or masses) and also in the rotational velocities. In order to take into account the intrinsic dispersion of the Tully-Fisher relation, we have added a dispersion factor ($\epsilon$) to the uncertainties of the K-band, stellar and baryonic masses, as suggested by Tremaine et al. (2002). The slope and zeropoints of the Tully-Fisher relation were obtained by using the IDL routine MPFITEXY \footnote{Torres-Flores (2011) used the FITEXY routine to obtain the different estimations of $\epsilon$ for the GHASP sample. In this work, we have compared the results obtained from FITEXY and MPFITEXY on the HCG sample and we found that both codes give the same values for $\epsilon$.} (Williams, Bureau \& Cappellari 2010), that depends on the MPFIT package (Markwardt 2009). Here, $\epsilon$ was automatically estimated in order to reach a $\chi^{2}$ of unity per degree of freedom. We note that in all the plots regarding the Tully-Fisher relation we have used the whole GHASP sub-sample studied by Torres-Flores et al. (2011), i. e. galaxies having baryonic masses between $8.91\leq\textrm{log}(M_{baryonic}/M_{\odot})\leq11.37$. This range in masses is larger than the range covered by the HCG sample studied in this work.

\subsection{Archival data} 

In order to extend our analysis to other wavelengths, we have searched for ancillary data. Photometry in
the u, g, r, i, and z bands and spectra were found in the \textit{SDSS} database, \textit{GALEX} NUV and FUV magnitudes were available in the \textit{GALEX} archive and photometric data points in the B and R-bands were taken from the NED database.

\section{Galaxy masses}

\subsection{Stellar masses}

In order to estimate the stellar mass of the galaxies, we need to assume a mass-to-light ratio ($\Upsilon_{\star}$). Several authors have used stellar population synthesis models to correlate $\Upsilon_{\star}$ with the colors of the galaxies (e.g. Bell \& de Jong 2001, Bell et al. 2003, Portinari et al. 2004). Recently, Torres-Flores et al. (2011) showed that the use of different models in the determination of the stellar masses does not change significantly the slope of the baryonic TF for the GHASP sample. These authors favor the recipes given by Bell et al. (2003) given that these models are an updated version of the models shown in Bell \& de Jong (2001) and given that these models have been used by other authors in the study of the stellar Tully-Fisher relation, which facilitate the comparison of the slope of this relation. In order to do a fair comparison of the HCG sample with the control sample, we have used the recipes given in Bell et al. (2003) to estimate the $\Upsilon_{\star}$ and the stellar masses of compact group galaxies (as also used in Torres-Flores et al. 2011 for the GHASP sample), following equation \ref{eq1}. In this case, we use the colors B-R given in Hickson (1993).

\begin{equation}
M_{\star}=10^{-0.264+0.138(B-R)} L_{K} 
\label{eq1}
\end{equation}

The K-band is susceptible to uncertainties in the contribution of TP-AGB stars (e.g. Portinari et al. 2004, Zibetti et al. 2009), which are not accounted for in the Bell et al. (2003) models. We then would expect that galaxies that have significant contributions from TP-AGB stars will stand out in the TF relations.

\subsection{Baryonic masses}

The mass of a galaxy is constituted of stars, HI gas, molecular gas and dark matter. The sum of the different gas masses to the stellar content corresponds to the baryonic matter. We have searched, in the literature, for individual HI measurements for the galaxies in our sample of compact group galaxies. Recently, Borthakur et al. (2010) presented the HI content of 22 Hickson compact groups, however, these authors did not list the HI masses of individual galaxies. On the other hand, Verdes-Montenegro et al. (2001) listed the individual HI masses for 13 galaxies in common with our sample: HCG 2a, 2b, 2c, 16a, 16c, 40c, 40e, 88a, 88b, 88c, 88d, 96a and 96c. When no HI measurement is available in the literature (for the other 22 galaxies) we computed the expected HI mass for each galaxy given its B-band luminosity and morphological type following the procedure given by Haynes \& Giovanelli (1984): 

\begin{equation}
M_{HI}=10^{c_{1}+c_{2}\times SM}\times L_{B}
\label{eqpredhi} 
\end{equation}

where c1, c2 and SM are values that depend on the morphological types and
magnitudes of the galaxies, respectively. Morphological types were taken
from Hickson (1993) and converted into the morphological system defined by Haynes \& Giovanelli (1984). B-band luminosities were computed using equation
\ref{lb}. In the case of HCG galaxies, it is well known that some of the
HI gas is found outside the disk of the galaxies, due to tidal effects. This may be an issue given that these environments are expected to have a high frequency of interaction. Interestingly, Verdes-Montenegro et al. (2001) noted that there is no one-to-one correlation between the presence of gaseous tidal tails and the HI content of compact groups, i. e. groups having tidal tails are not necessarily deficient in HI. Verdes-Montenegro et al. (2001) found that HCG galaxies have, on average, about 24$\%$ of their expected HI masses. Thus the computed HI masses (using equation \ref{eqpredhi}) for HCG galaxies may be upper limits. 

In an attempt to refine our estimation for the individual HI masses of HCG galaxies (with no HI observations), we have scaled the individual predicted HI masses depending on their morphological types. To do this exercise, we have used the sample of HCG galaxies for which Verdes-Montenegro et al. (2001) published the observed and predicted HI masses (Table 3 in that article). We divided that sample (37 galaxies) in four bins of morphological types (T) and then we estimated the fraction of the observed HI mass with respect to the predicted HI mass (for each morphological bin). We found that galaxies with morphological types between 10$\geq$T$\geq$7, 7$>$T$\geq$4, 4$>$T$\geq$0 and 0$>$T$\geq$-5 have 49\%, 81\%, 26\% and 19\% of their predicted HI masses, respectively. Therefore, we used these scaling factors to correct the predicted HI masses of the HCG for which there are no observations (and were the masses were estimated by using equation \ref{eqpredhi}). Errors in the predicted-and-scaled HI masses were taken as the mean of the difference between the predicted and observed HI masses for each morphological bin. The same errors were adopted for the observed HI masses (given that no errors were quoted by Verdes-Montenegro et al. 2001).

Torres-Flores et al. (2011) showed that the inclusion of the H$_{2}$
gas component in the computation of the baryonic masses of GHASP galaxies
does not affect significantly the slope of the baryonic TF relation,
therefore, we have not taken this component into account in this study.

The total mass in gas is related to the HI mass through: 

\begin{equation}
M_{gas}=1.4M_{HI}
\end{equation}

(McGaugh et al. 2000), where the factor 1.4 takes into account the correction for helium and metals. Finally, the baryonic mass is defined as the sum of the stellar and gaseous content:

\begin{equation}
M_{bar}=M_\star+M_{gas} 
\end{equation}

\begin{figure*}
\includegraphics[scale=0.55]{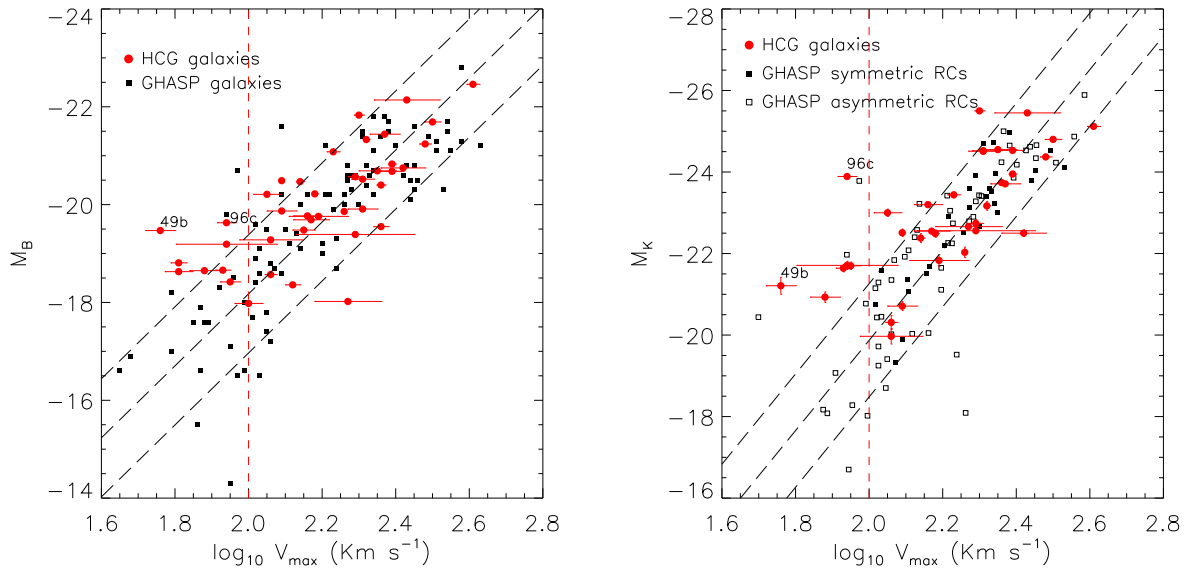}
\caption{B-band (left panel) and K-band (right panel) Tully--Fisher relation for the HCG sample (red filled dots). In the left panel, the GHASP sample is plotted as black filled squares. In the right panel, we have divided the GHASP sample for galaxies that display symmetric rotation curves (filled black squares) and asymmetric rotation curves (empty squares), following Torres-Flores et al. (2011). The black dashed lines represent the linear fit and a dispersion of 1$\sigma$ (see \S2.6) on the GHASP data. The red vertical line indicate a rotational velocity of 100 km s$^{-1}$.}
\label{hcgplots_MK}
\end{figure*}

\begin{figure*}
\includegraphics[width=\columnwidth]{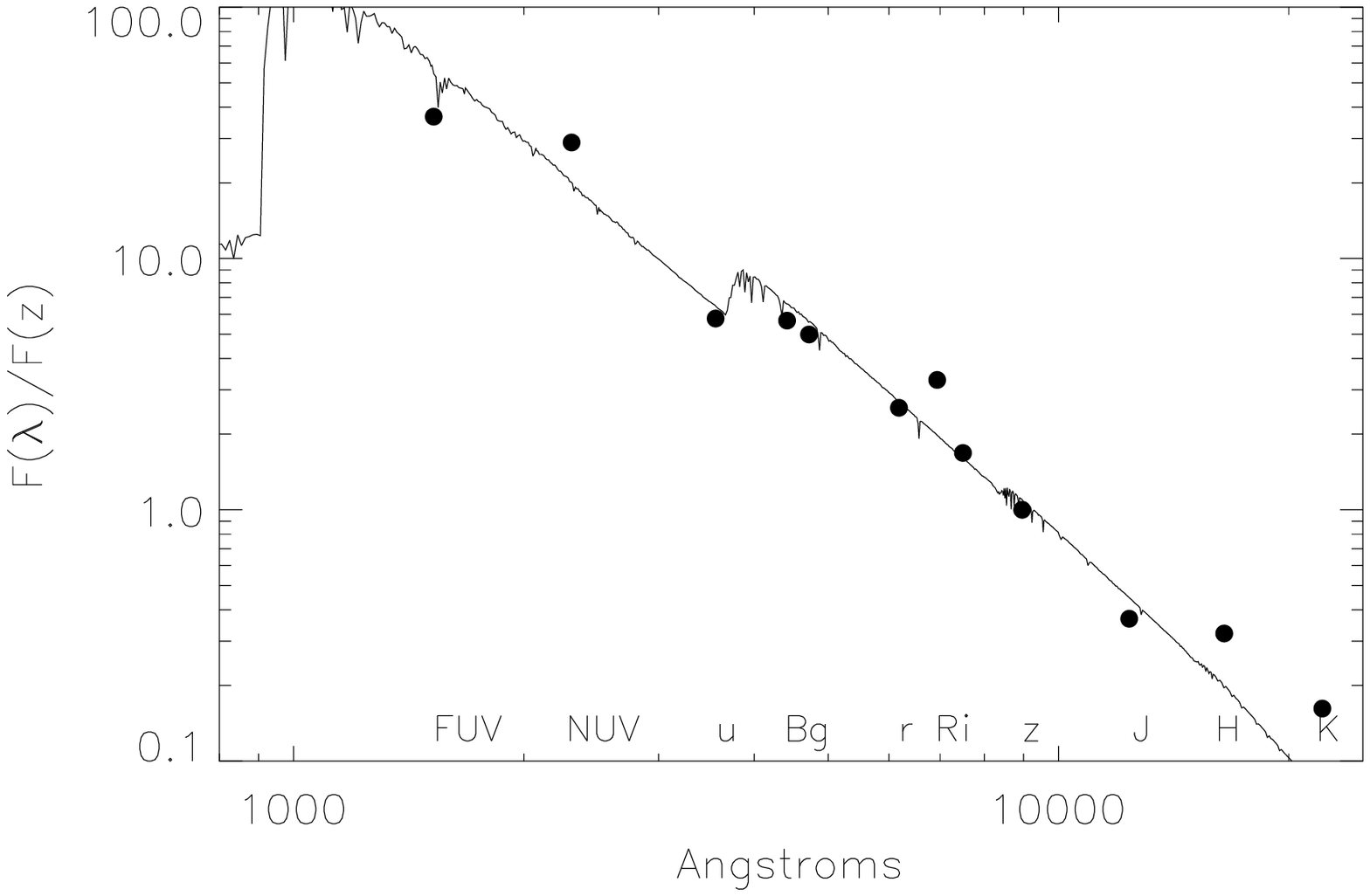}
\includegraphics[width=\columnwidth]{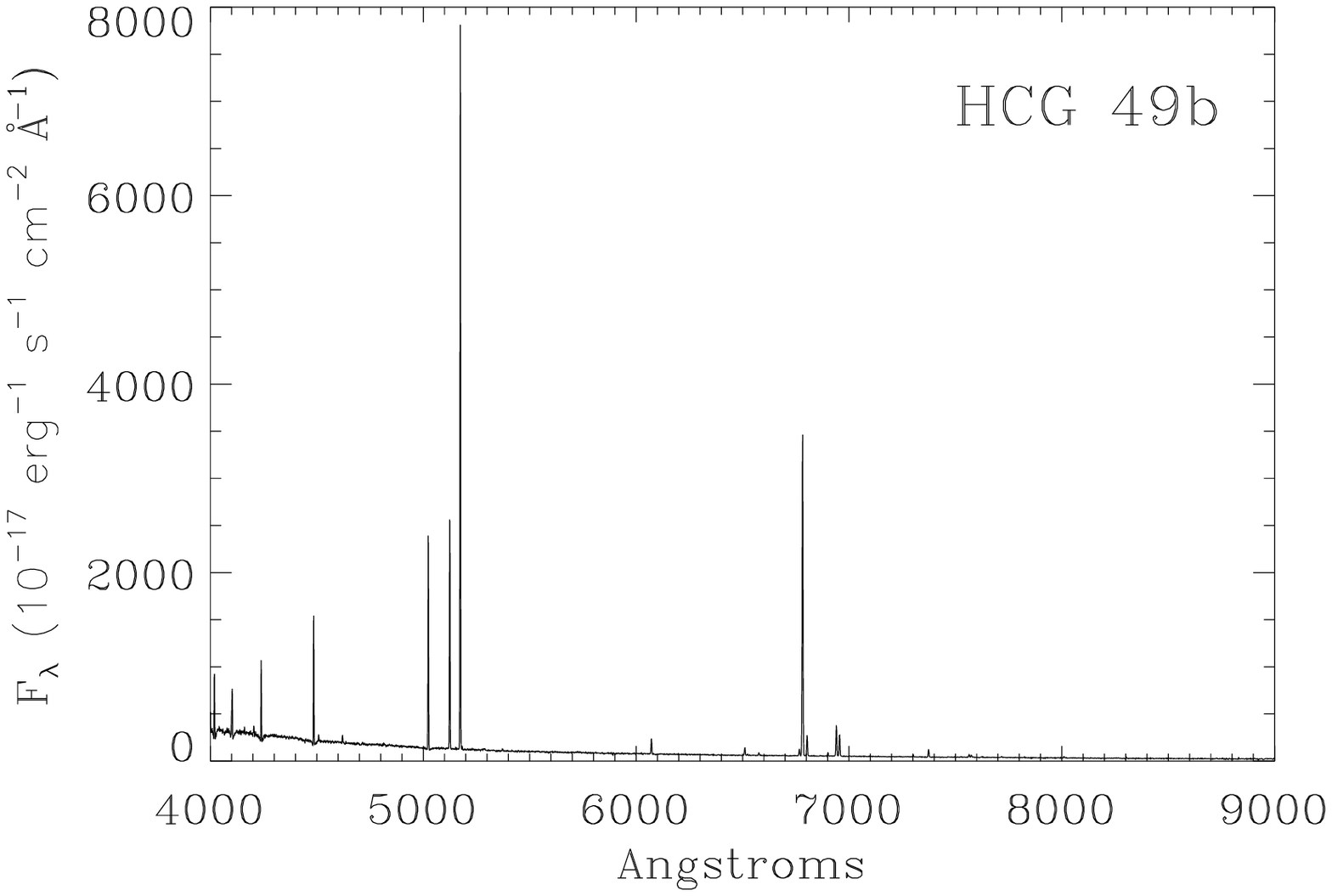}
\caption{Left Panel: Spectral energy distribution for HCG 49b. The continuous line corresponds to a single stellar population model, with an age of 8 Myrs, Z=Z$_{\odot}$/5 and Salpeter IMF. Filled dots represent the observed fluxes. Right panel: Optical spectrum of HCG 49b (taken from \textit{SDSS}), which shows the absence of continuum emission in this galaxy.}
\label{spectra_49b}
\end{figure*}

\begin{table*}
\centering
\begin{minipage}[t]{\textwidth}
\caption{Main parameters of the HCG sample}
\centering
\begin{tabular}{cccccccccc}
\hline
\multicolumn{1}{c}{Galaxy}& M${_K}$ & V$_{max}$ & B-R &$\Upsilon_{K}$ & log M$_{\star}$ & log M$_{gas}$ & log M$_{bar}$ & log M$_{tot}$\\
\multicolumn{1}{c}{HCG} & mag &km s$^{-1}$ & mag & & M$_\odot$ &M$_\odot$ & M$_\odot$ &M$_\odot$ \\
\multicolumn{1}{c}{(1)}&(2)&(3)&(4)&(5)&(6)&(7)&(8) &(9)\footnotetext{Column (1): Galaxy identification. Column (2): K-band absolute magnitude. Note that the value for galaxy 2a 
(marked with an asterisk) was obtained by us, when reanalyzing the 2MASS K-band image, and it is 0.77 mag brighter than the value 
given in 2MASS (m$_{K20}$=12.17 instead of 11.40 mag). Column (3): Maximum rotational velocity. V$_{max}$ values were taken from 
Torres-Flores et al. (2010). Column (4): Colors B-R taken from Hickson (1993). Columns (5): mass-to-light ratio calculated from 
Bell et al. (2003). Columns (6): stellar masses estimated by using $\Upsilon_{K}$ (column 5) and the K-band luminosity of each 
galaxy, except in the case of HCG 49b and HCG 96c (see \S4.2). Column (7): Gaseous masses. These values correspond to the HI masses corrected by helium and metals, i. e. $M_{gas}=1.4M_{HI}$. 
In the case of HCG 2a, 2b, 2c, 16a, 16c, 40c, 40e, 88a, 88b, 88c, 88d, 96a and 96c we used the observed HI masses listed in 
Verdes-Montenegro et al. (2001). In the remaining cases we used the predicted-scaled HI masses (see \S2.5). 
Column (8): Baryonic masses (M$_{\star}$+M$_{gas}$). Column (9): Total masses at the optical radius R$_{25}$ (estimated as $M_{tot} = V_{max}^{2}R_{25}/G$)}\\
 \hline
2 a & -22.50$\pm$0.08*  &  264$\pm$39 & 0.88& 0.72    &10.22$\pm$0.11  & 10.33$\pm$0.14 & 10.58$\pm$0.09& 11.27    \\
2 b & -22.56$\pm$0.07  & 196$\pm$74  & 1.14 & 0.78    &10.28$\pm$0.10  &  9.34$\pm$0.69 & 10.33$\pm$0.12& 10.72   \\
2 c & -20.71$\pm$0.11  & 122$\pm$13  & 1.10 & 0.77    & 9.54$\pm$0.11  &  9.70$\pm$0.31 &  9.92$\pm$0.19& 10.51    \\
7 a & -24.55$\pm$0.02  &  226$\pm$27 & 1.48 & 0.87    &11.12$\pm$0.10  &  9.47$\pm$0.74 & 11.13$\pm$0.10& 11.25   \\
7 c & -23.44$\pm$0.07  &  168$\pm$7  & 1.19  & 0.79   &10.64$\pm$0.10  &  9.89$\pm$0.20 & 10.71$\pm$0.09& 11.00   \\
7 d & -21.64$\pm$0.09  &  85$\pm$4   & 1.12 & 0.78    & 9.91$\pm$0.11  &  8.81$\pm$2.29 &  9.94$\pm$0.20& 10.23    \\
10 d &-22.55$\pm$0.06 &  149$\pm$16 & 1.52  & 0.88    &10.33$\pm$0.10  &  9.05$\pm$1.94 & 10.35$\pm$0.14& 10.66    \\
16 a &-24.53$\pm$0.02  & 247$\pm$17 & 0.82 & 0.71     &11.03$\pm$0.10  &  9.22$\pm$1.34 & 11.03$\pm$0.10& 11.08 	 \\
16 c &-23.74$\pm$0.03 &  229$\pm$6  & 1.04  & 0.76    &10.74$\pm$0.10  &  9.63$\pm$0.35 & 10.77$\pm$0.10& 10.97    \\
19 a &-23.20$\pm$0.04  & 144$\pm$13  & 1.56& 0.89     &10.60$\pm$0.10  &  8.46$\pm$5.21 & 10.60$\pm$0.11& 10.53 	 \\
19 b &-21.70$\pm$0.09  &  90$\pm$5   & 1.28& 0.82     & 9.96$\pm$0.11  &  8.72$\pm$4.17 &  9.98$\pm$0.25& 10.07    \\
19 c &-19.97$\pm$0.19  &  115$\pm$22 & 1.12& 0.78     & 9.24$\pm$0.13  &  9.30$\pm$1.51 &  9.57$\pm$0.81& 10.42    \\
37 d &-20.93$\pm$0.13  & 76$\pm$7    & 0.51& 0.64     & 9.54$\pm$0.11  &  8.83$\pm$4.44 &  9.62$\pm$0.73&  9.81    \\
40 c &-24.51$\pm$0.03  & 203$\pm$3   & 2.00& 1.03     &11.18$\pm$0.10  &  9.11$\pm$1.72 & 11.18$\pm$0.10& 11.17 	 \\
40 e &-22.66$\pm$0.07  & 186$\pm$40  & 1.84& 0.98     &10.42$\pm$0.10  &  8.37$\pm$9.47 & 10.42$\pm$0.13& 10.77 	 \\
49 a &-21.83$\pm$0.15 & 154$\pm$28  & 1.04  & 0.76    & 9.98$\pm$0.12  &  9.60$\pm$0.38 & 10.13$\pm$0.14& 10.72    \\
49 b &-21.21$\pm$0.21 & 58$\pm$6    & 0.90  & 0.72    & 8.00  &  9.31$\pm$1.49 &  9.33$\pm$1.42&  9.83    \\
56 a &-24.52$\pm$0.08 & 205$\pm$14  & 1.51  & 0.88    &11.12$\pm$0.11  &  9.87$\pm$0.21 & 11.14$\pm$0.10& 11.17    \\
68 c &-23.95$\pm$0.03  & 244$\pm$8   & 1.10& 0.77     &10.83$\pm$0.10  &  9.39$\pm$0.89 & 10.85$\pm$0.10& 11.25        \\
87 a &-25.12$\pm$0.04  &  410$\pm$16 & 1.86& 0.98     &11.40$\pm$0.10  &  9.67$\pm$0.32 & 11.41$\pm$0.10& 11.92        \\
87 c &-22.74$\pm$0.10  &  195$\pm$9  & 1.28& 0.82     &10.37$\pm$0.11  &  9.85$\pm$0.21 & 10.49$\pm$0.10& 11.04        \\
88 a &-24.80$\pm$0.04  &  315$\pm$16 & 1.56& 0.89     &11.24$\pm$0.10  &  9.27$\pm$1.19 & 11.24$\pm$0.10& 11.61 \\
88 b &-24.37$\pm$0.05  &  303$\pm$11 & 1.51& 0.88     &11.06$\pm$0.10  &  9.66$\pm$0.49 & 11.07$\pm$0.10& 11.45 \\
88 c &-22.51$\pm$0.09  &  123$\pm$2  & 0.99& 0.75     &10.24$\pm$0.11  & 10.16$\pm$0.11 & 10.50$\pm$0.08& 10.58 \\
88 d &-22.49$\pm$0.09  &  153$\pm$3  & 1.31& 0.83     &10.28$\pm$0.11  &  9.73$\pm$0.29 & 10.38$\pm$0.10& 10.84 \\
89 a &-23.17$\pm$0.11  &  210$\pm$4  & 1.16& 0.79     &10.53$\pm$0.11  &  9.98$\pm$0.16 & 10.64$\pm$0.09& 11.23 \\
89 b &-22.38$\pm$0.11  &  138$\pm$3  & 1.01& 0.75     &10.19$\pm$0.11  &  9.69$\pm$0.31 & 10.31$\pm$0.11& 10.72 \\
89 c &-22-03$\pm$0.13  &  180$\pm$1  & 1.11& 0.77     &10.07$\pm$0.11  &  9.54$\pm$0.44 & 10.18$\pm$0.13& 10.80 \\
91 a &-25.45$\pm$0.04	&  271$\pm$55 & 1.09& 0.77    &11.43$\pm$0.10  & 10.15$\pm$0.11 & 11.45$\pm$0.10& 11.52 \\
91 c &-23.00$\pm$0.09	&  112$\pm$10 & 1.08& 0.77    &10.45$\pm$0.11  &  9.60$\pm$0.38 & 10.51$\pm$0.11& 10.51 \\
93 b &-23.71$\pm$0.05  & 235$\pm$22  & 1.30& 0.82     &10.76$\pm$0.10  & 10.07$\pm$0.13 & 10.84$\pm$0.09& 11.35 \\
96 a &-25.50$\pm$0.05	&  201$\pm$6  & 1.33& 0.83    &11.48$\pm$0.10  & 10.06$\pm$0.19 & 11.50$\pm$0.10& 11.25 \\
96 c &-23.89$\pm$0.06	&  88$\pm$5   & 1.61& 0.91    &9.65  &  9.28$\pm$1.60 & 9.81$\pm$0.47& 10.08 \\
100 c&-21.71$\pm$0.10  &  87$\pm$27  & 1.23& 0.80     & 9.95$\pm$0.11  &  9.32$\pm$0.73 & 10.04$\pm$0.16& 10.14    \\
100 d&-20.31$\pm$0.17  &  114$\pm$4  & 1.18& 0.79     & 9.39$\pm$0.12  &  9.20$\pm$0.97 &  9.60$\pm$0.39& 10.25    \\
\hline
\label{table1}
\end{tabular}
\end{minipage}
\end{table*}

In Table \ref{table1} we list the main parameters of each HCG galaxy. Columns 1 to 4
give the name, K-band absolute magnitudes, maximum rotational velocities
and B-R colors. Column 5 lists the mass-to-light ratio for each galaxy, following Bell et al. (2003). Values for the stellar
masses are shown in column 6. In column 7 we list the total gas masses, calculated using
the observed HI masses (for HCG 2a, 2b, 2c, 16a, 16c, 40c, 40e, 88a, 88b, 88c, 88d, 96a and 96c) and the predicted and corrected/scaled HI masses (for the remaining galaxies), as described in \S2.5. In both cases the HI masses were corrected by helium and metals. Column 8 corresponds to the baryonic masses (M$_{\star}$+M$_{gas}$). Finally, in column 9 we list the total mass for each galaxy at the optical radius R$_{25}$ (estimated as $M_{tot} = V_{max}^{2}R_{25}/G$).

\section{Tully-Fisher relations}

\subsection{K and B-band Tully-Fisher relations}

The K-band TF relations are plotted for HCGs (red filled circles) and GHASP galaxies (black squares) in the right panel of Fig. \ref{hcgplots_MK}, where the filled black squares represent symmetric rotation curves while empty black squares show asymmetric curves. To allow a comparison of the
K-band TF relation with the B-band relation we included in the left panel
of Fig. \ref{hcgplots_MK} the corresponding B-band TF relation for the same
samples (Epinat et al. 2008b and Torres-Flores et al. 2010). Given that neither Torres-Flores et al. (2010) nor Epinat et al. (2008b) published the errors in the B-band magnitudes, the black dashed lines of the B-band Tully-Fisher relation were estimated from an ordinary least square bisector fit on the GHASP data. In the case of the K-band relation, a linear fit for the HCG galaxies yields: 

\begin{equation}
M_{K}=(-10.43\pm1.94)-(5.68\pm0.87)[log(V_{max})]
\end{equation}

Here we obtained an intrinsic dispersion factor of 0.97 mag. In the case of compact groups, most of the galaxies lie on the K-band TF
relation defined by the GHASP galaxies (black dashed lines), but presenting a larger dispersion than field galaxies (0.97 versus 0.76). Some of the low-mass galaxies show higher luminosities for their measured rotation velocities or too low rotational velocities for their luminosities. This
result was also found for the B-band TF relation studied in Mendes de
Oliveira et al. (2003) and Torres-Flores et al. (2010) as can be seen
in the left panel of Fig. \ref{hcgplots_MK}. More specifically, HCG 49b and HCG 96c are off
the relation defined by the GHASP galaxies (in the K-band) by more
than 2$\sigma$.

Interestingly, HCG 96c is located within 2$\sigma$ in the B-band TF
relation (left panel in Fig. \ref{hcgplots_MK}). In the case of HCG 49b,
this galaxy is outside the K-band and optical Tully-Fisher relations by
more than 2$\sigma$. Torres-Flores et al. (2011) found two
field galaxies from the GHASP sample that are also off the B and K-band Tully-Fisher relation
by more than 2$\sigma$ (having higher luminosities for their rotation or too low rotational velocities for their luminosities). However, we note that the GHASP galaxies that lie off the Tully-Fisher relation display asymmetric rotation curves and therefore large non-circular motions. Also, all but one GHASP outlier galaxy displays rising rotation curves which suggests that these objects probably do not reach their maximum rotational velocities (this is not the case of HCG 49b and HCG 96c).

\begin{figure*}
\includegraphics[scale=0.55]{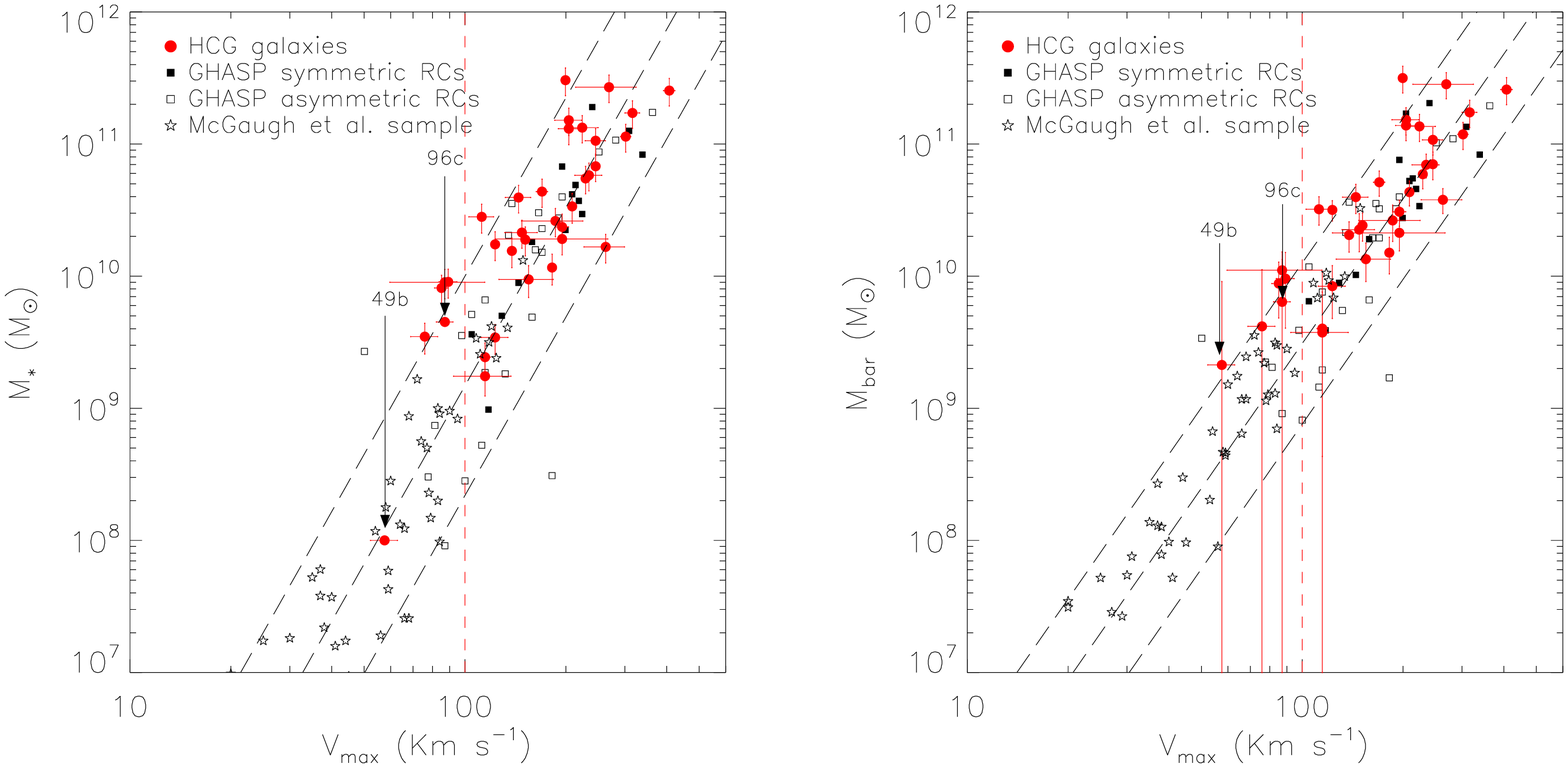}
\caption{Stellar (left panel) and baryonic (right panel) Tully--Fisher relation for the HCG sample (red filled dots). In both panels, we have divided the GHASP sample for galaxies that display symmetric rotation curves (filled black squares) and asymmetric rotation curves (empty squares), following Torres-Flores et al. (2011). The black dashed lines represent the linear fit and a dispersion of 1$\sigma$ (see \S2.6) on the GHASP data. Gas-rich galaxies (McGaugh et al. 2012) are shown by empty stars. The red vertical line indicate a rotational velocity of 100 km s$^{-1}$.}
\label{hcgplots}
\end{figure*}

\subsection{Stellar TF relation}

Stellar masses were at first obtained for all galaxies in our sample using equation 4. Given that we are using absolute magnitudes of HCG galaxies to derive their stellar masses we may expect that the outliers from the K-band Tully-Fisher relation, HCG 49b and HCG 96c, may display high stellar masses for their rotational velocities, if we use equation 4, which is indeed the case.

This gave us a hint that for these two galaxies, for some reason (described below and fully discussed in \S 4), we may need a more representative determination of the stellar mass, independent of the luminosity of the galaxy.

This was indeed confirmed for HCG 49b from its spectrum shown in the right panel of Fig. \ref{spectra_49b}, typical of an HII galaxy. Certainly equation 4 will not give a good determination for the stellar mass of a highly star bursting galaxy. Likewise, although we do not have an available spectrum, for HCG 96c Laurikainen \& Moles (1988) found that it has line ratios similar to those present in HII regions and they concluded that this galaxy could be fit with models with a recent burst of star formation. In the case of HCG 96c, we also had an additional hint that its K-band luminosity was not a good representation of the stellar mass given that the stellar mass obtained from equation 4 was in fact larger than that obtained from its rotation curve. Therefore, these two galaxies had their stellar masses computed differently from the rest of the sample, i. e. for these we did not use equation 4.

We describe below what was done in these two cases. We have searched for spectroscopic and multi wavelength photometric information in the literature with the aim of estimating not only their stellar masses but also other physical parameters (e. g. metallicities). These will be further discussed in \S 4.

\subsubsection{HCG 49b}

Using the spectral synthesis model STARBURST99 (hereafter SB99,
Leitherer et al. 1999) and observed fluxes taken from the literature,
we have fitted the spectral energy distribution (SED) of the galaxy
HCG 49b.  We have assumed an instantaneous burst, with a Salpeter Initial Mass Function (IMF)
and metallicities of Z$_{\odot}$=0.004, 0.008 and 0.02 (where the latter value correspond to the solar case). Observed fluxes range from far ultraviolet to near infrared,
using FUV and NUV from \textit{GALEX}, u, g, r ,i from \textit{SDSS},
B-band, R-band from NED  and J, H, K from the 2MASS database. 
We corrected the FUV flux by internal extinction (A$_{FUV}$), following the recipes given in Cortese
et al. (2008) and Seibert et al. (2005). We have used these two methods in order to know which one reproduces the true UV light for HCG 49b. Also, we used these recipes
because these authors used explicitly the FUV-NUV colors to estimate
the internal extinction instead of using far infrared data (60 and 100$\mu$m
fluxes), which are contaminated by flux from the neighbors, in the 
case of HCG 49b. For this galaxy, we obtain A$_{FUV}$=1.73 and A$_{FUV}$=2.95, following
Cortese et al. (2008) and Seibert et al. (2005), respectively. In order
to correct for internal extinction of HCG 49b in other bands, we have
used the A$_{FUV}$ value and the recipe from Boselli et al. (2003),
using the galactic extinction law given in Savage \& Mathis
(1979). In Fig. \ref{spectra_49b} (left panel) we show the result of the spectral energy
distribution fitting. The best fit was obtained using a $\chi^{2}$ test,
for a burst of 8 Myr, Z=Z$_{\odot}$/5, a initial mass of 10$^{8}$ M$_{\odot}$ and a FUV magnitude corrected by internal dust following Cortese et al. (2008). 

In order to determine the oxygen abundance of HCG 49b, we have used its optical spectrum, taken from the \textit{SDSS}, which is shown in Fig. \ref{spectra_49b} (right panel). Reddening was corrected using the Balmer lines ratio. Intrinsic H$\alpha$/H$\beta$ ratio was taken from Brocklehurst (1971) for an effective temperature of 10000 K and N$_{e}$=100. In order to estimate the metallicity of HCG 49b, we have used the N2 calibrator, proposed by Denicol\'o, Terlevich, \& Terlevich (2002), which is defined as the logarithm of the [N II] 6584/H$\alpha$ ratio. The range in which this ratio can be used as a proxy for metallicity estimation corresponds to $-2.6<\textrm{N2}<0.0$ (Denicol\'o, Terlevich, \& Terlevich 2002, see also Fig. 4 in Queyrel et al. 2011) which is the range in which HCG 49b lies. The oxygen abundance derived for this galaxy, using the \textit{SDSS} spectrum was 12+log(O/H)=8.24$\pm$0.14. Using the log[NII]/H$\alpha$ ratio given in Mart\'inez et al. (2010) for HCG 49b, we found a similar oxygen abundance of 12+log(O/H)=8.15$\pm$0.14. These values are typical of moderately low metallicity galaxies, given that solar metallicity is 12+log(O/H)$_{\odot}$=8.91 (Denicol\'o, Terlevich, \& Terlevich 2002) and they are consistent with the sub-solar abundance derived from the SED analysis.

We also computed the H$\alpha$ luminosity for an instantaneous burst and a sub
solar metallicity (Z$_{\odot}$=0.004). We compared this result with the
observed H$\alpha$ luminosity given in Mart\'inez et al. (2010). We
found that the observed H$\alpha$ luminosity corresponds to a burst
of $\sim$9 Myrs.  This value for the age is, however, an upper limit,
given that the observed luminosity is not corrected by dust. This age
is in agreement with the value obtained from the SED fitting analysis shown above.

\subsubsection{HCG 96c}

There was no spectrum available in the \textit{SDSS} for the galaxy HCG 96c. We then used the log[NII]/H$\alpha$ ratio given in Mart\'inez et al. (2010) to estimate an oxygen abundance of 12+log(O/H)=8.93$\pm$0.08 for this galaxy. This result is discussed in \S 4.2.2. For HCG 96c, Plana et al. (2010) studied its mass distribution. In the case of the stellar disk of this object, these authors found a mass of 4.5$\times$10$^{9}$ M$_{\odot}$ and 3.1$\times$10$^{9}$ M$_{\odot}$, when an isothermal sphere and a NFW model were assumed for the dark halo distribution, respectively. Given that the isothermal sphere model gives the highest value for the stellar disk of HCG 96c (and therefore an upper limit to the stellar mass), we will use this value as the total stellar mass of this object (Plana et al. 2010 did not published the uncertainties in the stellar masses).

\subsubsection{Final Stellar Tully-Fisher relation}

In Fig. \ref{hcgplots} (left panel) we plot the stellar Tully-Fisher relation for the HCG sample (red filled circles). The GHASP galaxies are represented by black filled and empty squares (as discussed in the previous section), while the McGaugh et al. (2012) sample is shown by empty stars. The black dashed lines correspond to the relation defined by the GHASP sample (Torres-Flores et al. 2011). In this Figure, two vertical arrows indicate the position of the galaxies HCG 49b and 96c. Inspecting Fig. \ref{hcgplots} (left panel) we found that most of the HCG galaxies lie on the stellar TF relation defined for galaxies in less dense environments. Four HCG galaxies lie slightly off the 1$\sigma$ relation defined by GHASP. Interestingly, the K-band Tully-Fisher outliers (HCG 49b and 96c) lie on the relation defined by GHASP, when their actual stellar masses are used. Specially in the case of HCG 49b, this object lies at the location where most of the McGaugh (2012) galaxies lie. In order to quantify the stellar TF relation for HCG galaxies, we have performed a linear fit on the data (equation \ref{stellar}). In this case, we obtained an intrinsic dispersion factor of 0.37 dex in the stellar mass.

\begin{equation}
M_{\star}=10^{(3.25\pm0.79)}V_{max}^{(3.22\pm0.36)}
\label{stellar}
\end{equation}

We find that the stellar TF relation for the HCG sample presents a larger dispersion 
than the relation defined for the GHASP sample, which has a dispersion factor of 0.31 dex in the stellar mass. We also find that the slope ($\alpha$) of the HCG TF relation differs significantly from the slope of the GHASP sample (which is $\alpha=4.38\pm0.38$). Torres-Flores et al. (2010) showed that the inclusion of HCG galaxies with rotational velocities lower than 100 km s$^{-1}$ in the computation of the B-band TF relation strongly affects the slope of this relation. In the case of the stellar relation, the exclusion of galaxies with rotational velocities lower than 100 km s$^{-1}$ (red vertical dashed line in Fig. \ref{hcgplots}) yields a slope of $\alpha=3.26\pm0.50$, which is similar to the slope derive for the whole HCG sample. Clearly, more low-mass HCG galaxies have to be observed before this matter ca be settled.

\begin{figure}
\includegraphics[scale=0.55]{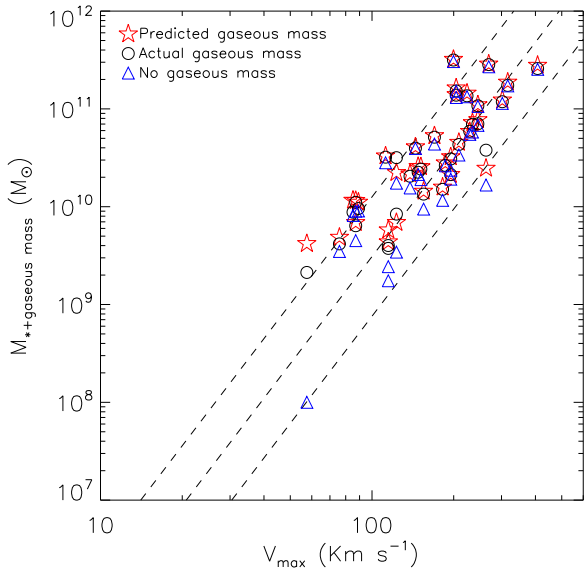}
\caption{Comparison between different assumptions of the HI mass in the baryonic TF relation. Predicted HI (and gaseous) mass: red stars. Actual gaseous mass: black circles. No gaseous mass: blue triangles. The baryonic Tully-Fisher relation defined by the GHASP sub-sample is indicated by dashed lines (Torres-Flores et al. 2011).}
\label{higashick}
\end{figure}

\subsection{Baryonic TF relation}

We plot in Fig. \ref{hcgplots} (right panel) the baryonic masses
obtained as described in \S 2.4 versus the rotational velocities for
each galaxy in the HCG (red filled circles) and GHASP (black filled and empty squares) samples. In the same figure, we have included the sample of McGaugh et al. (2012, empty stars). In this baryonic TF relation we find that most of the HCG galaxies lie on the relation defined for galaxies in less dense environments (black dashed lines, Torres-Flores et al. 2011). Taking into account the error bars, just two HCG galaxies lie off the 1$\sigma$ relation defined by the GHASP sample. Given the large uncertainties in the gaseous mass estimation, HCG low-mass galaxies display large error bars. We note that two GHASP galaxies lie off the 1$\sigma$ relation. Both galaxies display asymmetric and rising rotation curves, which suggests that these objects probably do not reach their maximum rotational velocity, which can explain their position on the TF relation (in fact, Torres-Flores et al. 2011 showed large uncertainties in the rotational velocity of these objects). Despite the few HCG galaxies which are slightly off the relation, it is clearly noticeable from a first inspection of the baryonic TF that this is the tightest relation among all TF relations studied so far in this paper (this point is discussed in more detail later on in this Section).

As explained in \S2.5, we have computed
a predicted-and-scaled HI mass for those HCG galaxies with no individual
measurements of HI available in the literature. To test if this or other
different assumptions (see below) will affect our results for
the baryonic TF relation, we plot in Fig. \ref{higashick} three slightly
different determinations of the mass using (1) the predicted gaseous mass,
with red stars, (2) the actual gaseous mass, with
 black circles (i.e. the individual gaseous mass
for the 13 galaxies in common with Verdes-Montenegro et al. 2001 and
the predicted-and-scaled HI masses for the remaining galaxies,
as described in \S2.5) and (3) no gaseous component, with blue triangles,
(i.e., we assumed no mass in the form of gas). Dashed lines represent the baryonic Tully-Fisher relation defined by the GHASP sub-sample (see bottom left panel of Fig. \ref{hcgplots}). From Fig. \ref{higashick}
we conclude that using the predicted or observed gaseous mass does
not influence strongly the position of the HCG galaxies in
this plot. 

In order to compare the parameters for the baryonic TF relation obtained
by using actual (i. e. observed for 13 objects and predicted-and-scaled 
HI masses for the remaining ones) and
predicted HI masses (from equation \ref{eqpredhi}), we used a linear fit (as described in \S 2.6) to quantify the slopes and zero points for the HCG sample (equations \ref{baryonicreal} and \ref{baryonicpre}).

\begin{equation}
M_{bar,actual HI}=10^{(5.20\pm0.69)}V_{max}^{(2.42\pm0.30)}
\label{baryonicreal}
\end{equation}

\begin{equation}
M_{bar,pred HI}=10^{(5.21\pm0.71)}V_{max}^{(2.41\pm0.31)}
\label{baryonicpre}
\end{equation}

In the first case, we obtained an intrinsic dispersion factor of 0.23 dex in the baryonic mass, while in the second case we found a dispersion factor of 0.24 dex (we note that these fits were applied on the logarithmic values of M$_{bar}$ and V$_{max}$). The fits listed above indicate that when the baryonic masses are used in the TF
relation (no matter if actual or predicted HI masses are used), the
uncertainties in the slope and zero point are smaller than in the case
of the stellar TF relation (equation \ref{stellar}). We also note that
the baryonic TF relation for the HCG galaxies (described by equations
\ref{baryonicreal} and \ref{baryonicpre}) is tighter than the stellar
relation shown in the left panel of Fig. \ref{hcgplots}. However, in
all cases, the uncertainties in these parameters are larger than the same
values derived for the GHASP sample (Torres-Flores et al. 2011). Also, the baryonic TF relation for HGC galaxies in our sample has a larger intrinsic dispersion factor (0.23 dex) than GHASP galaxies (which present a dispersion factor of  0.21 dex). We note that the use of the actual and/or predicted-and-scaled HI masses in the determination of the slope of the baryonic TF relation produces a shallower slope than the value derived for the GHASP
sample ($\alpha=3.64\pm0.28$).

\begin{figure}
\includegraphics[scale=0.55]{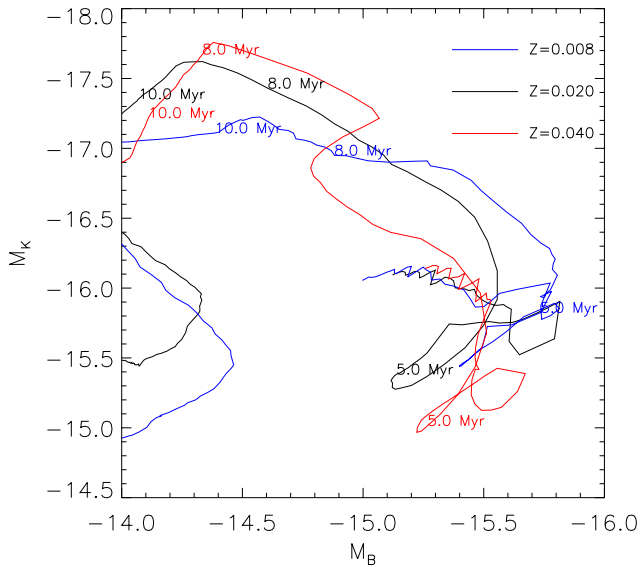}
\caption{Evolution of the B and K-band absolute magnitudes as a 
function of time for a single stellar population (SB99 models). Different ages are labeled in this figure. 
We assumed an instantaneous burst, with a Salpeter IMF.}
\label{mk_mb_49b}
\end{figure}

\section{Discussion} 

In this paper, we have computed masses of HCG galaxies and have compared the
K-band, stellar and baryonic TF relations for galaxies in two different
environments: in the field (GHASP sample) and in compact groups (HCG sample). As a complement, we have included the sample of gas-rich galaxies studied by McGaugh (2012). In the case of GHASP and HCG samples, the data were analyzed in the same way. Here, we summarize our main findings in
this study so far:

\begin{enumerate} 

\item We found that most of the HCG galaxies lie on the K-band Tully-Fisher relation defined by field galaxies (GHASP survey and even gas-rich galaxies) with a couple of exceptions for the galaxies HCG 49b and HCG 96c.

\item We fit the SED of HCG 49b using a single stellar population and we estimated its actual stellar mass, which corresponds to $\sim$10$^{8}$ M$_{\odot}$. Using this value and the already published stellar mass of HCG 96c, we found that these K-band Tully-Fisher outliers galaxies lie on the stellar and baryonic Tully-Fisher relations defined by field galaxies.

\item The scatter presented by HCG galaxies in the K-band, stellar and
baryonic TF relations is always higher than the scatter of corresponding
relations for field galaxies (GHASP sample).

\item The scatter presented by HCG galaxies in the baryonic TF relation
is smaller than that presented in the stellar TF relation, no matter if
predicted or observed HI masses are taken into account. This confirms
the importance of taking the gas component into account when determining
the mass of a galaxy.

\end{enumerate} 

\section{The case of the K-band outliers galaxies HCG 49b and HCG 96c}

Given the importance of the scatter in the Tully-Fisher relation, specially at high redshifts (where galaxy-galaxy interactions are quite common), we have studied the physical properties of the two K-band Tully-Fisher HCG outliers: HCG 49b and HCG 96c. By using that information we will discuss the possible explanations for the location of the outlying low-mass HCG galaxies in the TF relation.

\subsection{Their environment}

In order to understand the location of HCG 49b and HCG 96c on the K-band Tully-Fisher relation, it is necessary to know the environment in which these galaxies lie. Both HCG 49b and HCG 96c are located in groups which are in quite advanced stages of evolution. 

\subsubsection{HCG 49b} 

For HCG 49, the four main galaxies of the system are known to have disturbed morphologies (Mendes de Oliveira et al. 1992). This system has an HI halo that encompasses all the galaxies, having a single HI velocity gradient (Verdes-Montenegro et al. 2001). In the evolutionary scenario proposed by Verdes-Montenegro et al. (2001), HCG 49 is classified as an evolved group. In addition, Torres-Flores et al. (2010) found that HCG 49b has a perturbed velocity field. This fact suggests that this galaxy has already experienced some interaction with its companions.

\subsubsection{HCG 96c} 

The four galaxies of HCG 96 show signs of gravitational interaction and in particular HCG 96c is part of a close pair (HGC 96ac), which displays two faint and long stellar tails (Verdes-Montenegro et al. 1997). The HI distribution of this system is peculiar. HCG 96a presents two gaseous tidal tails that enclose the members HCG 96c and HCG 96d, suggesting that galaxy-galaxy interactions have taken place in this group (Verdes-Montenegro et al. 2000). Amram et al. (2003) found that HCG 96c has a peculiar velocity field, where the isovelocities are not regular. As in the case of HCG 49, this group has already experienced some events of galaxy-galaxy interactions.

\subsection{Enhancement in star formation and consequent brightening of 
the galaxies} 

Mendes de Oliveira et al. (2003) suggested that a few low-mass HCG
galaxies have enhanced B-band luminosities due to star formation, most
probably triggered by galaxy-galaxy interactions (where HCG 49b can be identified as an outlier). In the K-band TF relation there are also at least two low-mass outliers (HCG 49b and HCG 96c). There may also be other examples (most probably
HCG 89d, HCG 96d) which are outliers in the B-band TF but for which no
K-band magnitudes are available.  In order to explain the position of
these galaxies in the TF relations we first analyze the possibility of
the presence of strong ongoing star formation in these galaxies.

\subsubsection{HCG 49b} 

In Fig. \ref{spectra_49b} we show the SED and the \textit{SDSS} optical spectra of HCG 49b (left and right panels, respectively). This object is clearly dominated by
star formation, displaying little continuum. As described in \S3.4,
the oxygen abundance measurement obtained for 49b is well below solar
12+log(0/H)=8.15$\pm$0.14. We found fairly good agreement between the observed and
the synthetic spectral energy distribution of HCG 49b. Also, assuming
an instantaneous burst, this object is apparently very young, having a
total mass of about 10$^{8}$M$_{\odot}$. In view of these results, it is
difficult to know if there is contribution of an old stellar population,
given that the spectrum of HCG 49b is dominated by the star formation
process that is taking place in this galaxy. In fact, HCG 49b has the
highest H$\alpha$ luminosity in the sample of 200 HCG galaxies studied
by Mart\'inez et al. (2010), even higher than the H$\alpha$ luminosity
of the individual members of the merger system HCG 31. Therefore, HCG 49b has clearly a recent burst of star formation, which could increase its luminosity. 

\subsubsection{HCG 96c} 

As determined in \S3.2, the oxygen abundance of HCG 96c is
12+log(O/H)=8.93$\pm$0.08. This value is slightly higher than the solar oxygen
abundance given in Denicol\'o, Terlevich, \& Terlevich (2002). If we
assume a solar or super solar metallicity and using starburst99 models,
we find that for an instantaneous burst of a few million years, the
B-band magnitude does not change much while the K-band magnitude can
change up to one magnitude (as shown in Fig. \ref{mk_mb_49b}), supporting our observations.

In fact, in a study of Seyfert galaxies and their companions,
Laurikainen \& Moles (1988) found that HCG 96c has line ratios
similar to those present in HII regions, finding also an over solar
metallicity for this object. Using the observed H$\alpha$ luminosity,
these authors suggested that a star-forming process is going on in this
galaxy. They also concluded that the spectrum of this galaxy could
be fit with models with a recent burst of star formation. In view of
these results and Fig. \ref{mk_mb_49b}, it is possible that HCG 96c
has changed its K-band luminosity in almost one magnitude and, at the
same time, keeping its B-band luminosity with the same value as before
the burst (Fig. \ref{mk_mb_49b}). Maraston (1998) studied the evolution
of stellar populations using synthetic models which take into account
the contribution of thermally pulsing asymptotic giant branch (TP-AGB)
stars. She found that for an age of $\sim$700 Myrs, there is a large
contribution of TP-AGB stars in the K-band luminosity, which is not
taken into account in most models. If HCG 96c has experienced a
burst which has evolved in time, the K-band luminosity could be strongly
affected by TP-AGB stars, which seems to be the case in view of its
position in the Tully-Fisher relation.

\subsubsection{Is the environment the main responsible to explain the outlier galaxies?}

The two K-band Tully-Fisher outliers HCG 49b and HCG 96c are located in interacting groups. Both galaxies have the typical rotational velocities of late-type spiral galaxies, which exclude a mass truncation in these systems, and both galaxies present signatures of ongoing and recent star formation. Also, HCG 96c appears as a transition object. Bitsakis et al. (2011) have recently showed that late-type galaxies in dynamically ``young'' HCG have similar star formation properties as field galaxies. Also, these authors did not find any evidence of  enhanced AGN activity in the different evolutionary stages of the groups. Therefore, HCG 49b and HCG 96c do not present the same physical properties as the general sample of galaxies in HCG. 

Despite the small numbers of galaxies involved here, which hamper any strong conclusions, we can speculate that the environment plays a significant role in the location of low-mass galaxies in the Tully-Fisher relation. Interacting low-mass galaxies located in dense environments may not follow the same Tully-Fisher relation defined by non-interacting galaxies, given that their luminosities can be increased due to bursts of star-formation produced in interaction events, as was also pointed out by Kannappan et al. (2004) for a sample of galaxy pairs. This idea is supported by the fact that HCG 49b and 96c lie on the stellar TF relation and therefore their masses are typical of low-mass systems. Given the data available in this study, we can tentatively suggest that if there are low-mass outliers to the K-band Tully-Fisher relation, these objects may be interacting galaxies. 

\subsection{K-band luminosity enhanced by AGN activity} 

Another mechanism to increase the K-band luminosity without affecting the B-band luminosity could be the presence of an AGN. Riffel et al. (2009) studied the near-infrared spectral energy distribution of a sample of nine Seyfert 1 and fifteen Seyfert 2 galaxies, finding that Seyfert 1 galaxies need a dust component to explain their higher emission in the K-band with respect to Seyfert 2 galaxies. This emission could be associated with the torus of dust, which could be heated by the AGN. Interestingly, HCG 96c is a transition object as defined by Mart\'inez et al. (2010), which means that this galaxy has AGN signatures. This scenario could explain an increase in the K-band luminosity, placing this galaxy off the TF relation. Infrared spectra for HCG 96c could elucidate this issue. HCG 49b is not an AGN, therefore, we exclude this scenario as a possibility for this galaxy. 

We note that two other galaxies in the same mass range as HCG 49b are
89d and 96d. They are also outliers in the B-band TF but unfortunately
we do not have any information on their NIR magnitudes and therefore we
do not know their locations in the baryonic TF relation.

\section{Conclusions and summary} 

The Tully-Fisher relation was mainly used in the past to determine galaxy distances and to measure deviations from the cosmic flow (from Tully \& Fisher 1977 to e.g. Springob et al. 2009). Nowadays the Tuly-Fisher relation is used to test galaxy formation and evolution models (e.g. Dutton et al. 2007) and for connecting the disk to the dark halo masses (e.g. Dutton et al. 2010). Also, the scatter in the Tuly-Fisher relation provides useful constraints on the intrinsic differences between disk galaxies and on their evolution related to their environment. 

In this paper, to tackle the question of the role of the environment in galaxy evolution, we have studied for the first time the K-band, stellar and baryonic Tully-Fisher relations for a sample of galaxies in compact groups, where tidal interactions are common, and we compared them with a sample of galaxies in less dense environments (GHASP survey). To complement our reference data in low density environment, we have included the sample of gas-rich galaxies studied by McGaugh et al. (2012). The HCG and GHASP samples, both based on 2D Fabry-Perot data obtained with the same kind of instrument, were analysed using the same procedure in order to minimize the difference liked to data analysis.

From this study, we concluded that the behaviour of most compact group galaxies on the different Tully-Fisher relations (K-band, stellar and baryonic) does not fundamentally differs of those shown by field galaxies, however, the larger scatter on the plots and the presence of some outliers indicate subtle, but intrinsic, differences between compact group and field galaxies. These differences can be likely linked to transient evolutionary phenomena due to the dense environment of compact group of galaxies. This means that within the compact groups dominated by late-type galaxies we can observe galaxies having different degrees of kinematic perturbations, which are caused by it environment. The more dynamically perturbed galaxies are dominated by non-rotating motions and they do not lie in the Tully-Fisher relation. On the other hand, less perturbed galaxies are still dominated by the rotation of the disk and they are similar to field galaxies. Between these two boundaries one observes intermediate cases with different degrees of perturbations, i .e. not high enough to leave the Tully-Fisher relations defined by non-interacting galaxies, but high enough to increase the scatters in the relations. These findings confirm evolutionary scenarios of compact group.

In the following, we summarize our main findings:

\begin{enumerate}

\item We found that most HCG galaxies lie on the K-band, stellar and baryonic Tully-Fisher relation defined by the galaxies of GHASP, but having a larger scatter. 

\item The scatter in the stellar HCG Tully-Fisher relation is slightly reduced when the gaseous masses are added to the stellar ones, i. e. in the baryonic Tully-Fisher relation.

\item Two low-mass galaxies (HCG 49b and HCG 96c) are shifted from the expected K-band Tully-Fisher relation and HCG 49b is also shifted from the B-band relation. However, both galaxies lie on the stellar and baryonic Tully-Fisher relations. Given that these objects are actively forming stars, we speculate that their positions on the B and K-band Tully-Fisher relations are a result of an enhancement in their luminosities triggered by star-formation events and maybe also AGN activity in the case of HCG 96c. These high luminosities do not affect the total stellar mass of these systems, and that is the reason why galaxies HCG 49b and 96c lie on the stellar and baryonic Tully-Fisher relations.

\end{enumerate}

We plan in the future to study other low-mass HCG galaxies to check how they fit in the TF relation.

\section*{acknowledgments}

We thank S. Boissier for helpful and stimulating discussions about
the spectral energy distribution of HCG 49b. S. T-F acknowledges the financial support of the chilean agency CONICYT through the Programa de Inserci\'on de Capital Humano Avanzado en la Academia, under contract 7912010004 and also to FAPESP
(doctoral fellowship, under contract 2007/07973-3). S. T-F also acknowledges
the financial support of EGIDE through an Eiffel scholarship. CMdO acknowledges support from FAPESP
(2006/56213-9) and CNPq. HP acknowledges financial support of CNPq (201600/2009-9 and 471254/2008-8). HP thanks CNPq/CAPES for its financial support using the PROCAD project 552236/2011-0. We also acknowledge the use of the HyperLeda database
(http://leda.univ-lyon1.fr). \textit{GALEX} is a NASA Small Explorer, launched in 2003 April. We gratefully acknowledge NASA's support for construction, operation, and science analysis for the \textit{GALEX} mission, developed in cooperation with the Centre National d' Etudes Spatiales of France and the Korean Ministry of Science and Technology. This research has made use of the NASA/IPAC
Extragalactic Database (NED) which is operated by the Jet Propulsion
Laboratory, California Institute of Technology, under contract with the
National Aeronautics and Space Administration.

\label{lastpage}
\end{document}